\documentclass{JHEP3}
\usepackage{amsmath}
\usepackage{axodraw}
\usepackage{epsfig}

\newcommand{\order}{\mathcal O}

\title{ \bf Double fermionic 
contributions to the heavy-quark vacuum polarization}

\author{Micha{\l} Czakon$^{1,2}$~and~Thomas Schutzmeier$^{1}$\\\\
      $^1$~{Institut f\"ur Theoretische Physik und Astrophysik, 
      Universit\"at W\"urzburg, \\
      Am Hubland, D-97074 W\"urzburg, Germany} \\ \\
      $^2$~{Department of Field Theory and Particle Physics,
            Institute of Physics, \\
            University of Silesia, Uniwersytecka 4, PL-40007 Katowice,
            Poland} }

\abstract{ 
We compute the virtual $\order{(\alpha_s^3\,n_f^2)}$ corrections to the
heavy quark vector current correlator in terms of expansions in
the external momentum and as an exact numerical solution. As a
byproduct, the available high-energy expansion at the
three-loop level is extended.  }

\begin{document}
\setlength{\parindent}{0pt}

\section{Introduction}

Correlators of two currents, basic objects in Quantum Field Theory,
provide important information for both theoretical and
phenomenological applications. Depending on the Lorentz structure of
the current under consideration, different interesting observables are
directly related to these quantities, like the hadronic cross section in
electron-positron annihilation, $R(s)$, and decay rates of Z- and
Higgs-bosons. 

On this account current correlators are investigated thoroughly in
perturbation theory, where even high order calculations are possible.
Some applications, like the determination of the charm and bottom
quark masses through sum-rules carried out to four loops
in~\cite{Boughezal:2006px,Kuhn:2007vp}, corrections to the
$\rho$~parameter \cite{Boughezal:2006xk,Chetyrkin:2006bj} or the
recent evaluation of $\alpha_s$ from low energy data at the same level
of precision~\cite{Kuhn:2007tc}, require asymptotic expansions of
heavy quark correlators in the low and/or high energy limits.  Other
studies, however, necessitate the knowledge of the full external
momentum dependence $p^2$, for instance the determination of the fine
structure constant at the Z-boson mass scale,
$\alpha_{\text{em}}(M_Z)$~\cite{Jegerlehner:2006ju}.

Up to the three-loop level, all physically relevant correlators have
been computed including the full quark mass dependence in
~\cite{Chetyrkin:1996cf, Chetyrkin:1997mb,Chetyrkin:1998ix} by
deriving Pad\'e approximants from asymptotic expansions. At four-loop
accuracy, in the case of the vector current including one heavy quark,
the first two terms in the low energy series are currently
known~\cite{Boughezal:2006px, Chetyrkin:2006xg} and were obtained by
direct Taylor expansion of all propagator-type integrals and
subsequent reduction of tadpole diagrams.  In the high energy limit,
besides the leading massless
contribution~\cite{Chetyrkin:1979bj,Celmaster:1980ji,Dine:1979qh,
Gorishnii:1990vf, Surguladze:1990tg, Chetyrkin:1996ez} , the first two
mass correction terms of the absorptive part of the scalar, vector and
axial-vector current correlators are also
available~\cite{Chetyrkin:1996hm,Chetyrkin:2000zk}, partially even to
five loops~\cite{Baikov:2001aa, Baikov:2002va}.

Unfortunately, the techniques used to achieve these results are, due to
their huge computational complexity, not suitable for the computation of
higher order terms in the expansions. Thus, reconstructing the full
momentum dependence at $\order(\alpha_s^3)$ in an analogous
manner to the three-loop case demands a completely different method.
A promising and effective approach in this context is based on
differential equations, originally proposed
in~\cite{Caffo:1998du,Remiddi:1997ny}. Using this technique, we were
already able to compute the first 30 terms in the low energy expanded
polarization function at $\mathcal O(\alpha_s^2)$
in~\cite{Boughezal:2006uu}.  Recently, these results were confirmed
in~\cite{Maier:2007yn} utilizing the same approach.

The purpose of this paper is to exploit this method and demonstrate
its applicability up to the four-loop level. We compute
the vacuum polarization for low and high energies as expansions up to
$(p^2)^{30}$ and $(p^2)^{-15}$, respectively, and moreover the full momentum
dependence in numerical form in the Euclidean and Minkowskian region.
As new results, we extend the available information in the high energy domain at
$\order(\alpha_s^2)$ and provide the double fermionic
contributions to the vector current correlator at $\order(\alpha_s^3)$.

This work is organized as follows. In the next section we sum up the
relevant notation. Calculational methods are briefly presented in
Section 3, whereas the first terms of the expansions and the numerical
results are given in Section 4. Finally, we give our conclusions.

\section{Definitions}
Given the vector current $j^{\mu}(x) = \overline q_h(x)\gamma^{\mu}q_h(x)$ 
composed of the heavy quark field $q_h(x)$ with mass $m$,
the two point correlator is defined by
\begin{equation}
  \Pi^{\mu\nu}(p^2) = i\,\int \,dx e^{ixp} \langle0|Tj^{\mu}(x)
  j^{\nu}(0)|0\rangle
\end{equation}
where $p^{\mu}$ is the external momentum. A convenient representation
of the tensor $\Pi^{\mu\nu}(p^2)$ through the scalar vacuum polarization 
function $\Pi(p^2)$ is given by
\begin{equation}
  \Pi^{\mu\nu}(p^2) = (-p^2 g^{\mu\nu} + p^{\mu}p^{\nu}) \Pi(p^2) +
  p^{\mu} p^{\nu} \Pi_L(p^2).
\end{equation}
Transversality of the vector current correlator requires $\Pi_L(p^2) =
0$. Through the optical theorem the aforementioned hadronic ratio
$R(s)$ is related to the current correlator 

\begin{equation}
  R(s) = \frac{\sigma(e^+e^- \to
  \text{hadrons})}{\sigma(e^+e^-\to\mu^+\mu^-)} =
  12\pi\text{Im}\Pi(p^2 = s + i\epsilon).
\end{equation}

In the framework of perturbation theory, the polarization function can
be expanded in the strong coupling as
\begin{equation}
  \Pi(p^2) = Q_h^2 \frac{3}{16 \pi^2}\sum_{k\geq0} \left(\frac{\alpha_s(\mu)}{\pi}\right)^k
  \Pi^{(k)}(p^2).
\end{equation}

\begin{figure}
  \begin{center}
    \epsfig{file=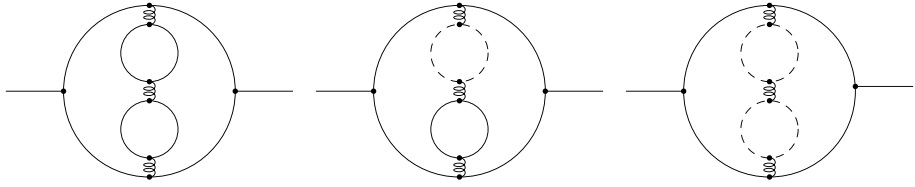, width=.8\textwidth}
  \end{center}
  \caption{Diagrams contributing to $\Pi^{(3)}_{\text{df}}(p^2)$. 
  Solid (dashed) lines refer to massive (massless) propagators.}
  \label{fig:diagrams}
\end{figure}

Since in this work the focus is set on double fermionic contributions
at $\order (\alpha_s^3)$ stemming form diagrams sketched in fig.
\ref{fig:diagrams}, it is convenient to decompose $\Pi^{(3)}(p^2)$
into bosonic and fermionic contributions,

\begin{eqnarray}
 \Pi^{(3)}(p^2) &=&  C_F^3\,\Pi^{(3)}_\text{A}(p^2)
  +C_F^2\,C_A\,\Pi^{(3)}_\text{NA,1}(p^2)
  +C_F\,C_A^2\,\Pi^{(3)}_\text{NA,2}(p^2) \\\nonumber&&
  +C_F^2\,T_F\,\Pi^{(3)}_\text{sf,A}(p^2)
  +C_F\,C_A\,T_F\,\Pi^{(3)}_\text{sf,NA}(p^2)\\\nonumber&&
  +C_F\,T_F^2\, \Pi^{(3)}_\text{df}(p^2),
\end{eqnarray}

where $\Pi^{(3)}_\text{sf}(p^2)$ and $\Pi^{(3)}_\text{df}(p^2)$ denote
terms proportional to $n_f$ and $n_f^2$, respectively, with $n_f$
being the number of active flavours. $C_F$ refers to the Casimir
operator and $T_F$ to the trace of the fundamental representation of
$SU(N)$. Because of the mass hierarchy in the quark sector only the
heavy quark is considered massive whereas all lighter ($n_l = n_f -
1$) quarks are treated as massless. 

\medskip

Using the fact that the double fermionic contribution contains no
massless cuts, $\Pi^{(3)}_\text{df}(p^2)$ can be expanded in a simple
power series in the low energy limit $p^2 \to 0$.  In the large energy
limit $p^2 \to -\infty$, however, non-integer powers of $p^2$ arise
and lead to additional logarithms of the form $\log(-p^2/m(\mu)^2)$.
Thus, defining $z=p^2/4m(\mu)^2$, we end up with the expansions

\begin{eqnarray}
  z \to 0: &\quad&
    \Pi^{(3)}_\text{df}(z) =  \sum_{n>0} C^0_n(\mu)
  \,z^n, \label{eq:lowzanzatz}\\
  z \to -\infty: &\quad&
    \Pi^{(3)}_\text{df}(z) =  \sum_{n,m} C^\infty_{nm}(\mu)\, z^{-n} \,\log^m(-z)\label{eq:largezanzatz}.
\end{eqnarray}

\section{Calculation}
The basic idea for computing the full $p^2$-dependence of the vacuum
polarization function is to deal with massive propagator-type
integrals. Although the number of integrals is moderate (approx.
$10^4$ for the whole four-loop contribution) they pose a challenge as
far as the reduction to a small set of master integrals is concerned,
since two variables, $z$ and $d=4-2\epsilon$, the dimension of
space-time, are involved.  For the determination of these master
integrals, however, an efficient method through differential equations
exists and allows for asymptotic expansions to high orders in the
external momentum and high precision numerics. 

\medskip

In a first step all Feynman diagrams contributing to the double
fermionic corrections have been projected onto scalar integrals, which
were subsequently reduced to a set of 46 master integrals with the
help of integration-by-parts (IBP) identities~\cite{Chetyrkin:1981qh}
and the Laporta algorithm~\cite{Laporta:2001dd} implemented in
\texttt{IdSolver}~\cite{IdSolver}. 

\medskip

To obtain the master integrals, we used the scaling property of
propagator-type integrals $P_i(p^2,m^2)$

\begin{equation}
  P_i(\lambda p^2, \lambda m^2) = \lambda^{D[P_i(p^2,m^2)]} P_i(p^2, m^2)
\end{equation}

to get the characteristic differential equation

\begin{equation}
  p^2 \frac{\partial}{\partial p^2} P_i(p^2,m^2)=-m^2\frac{\partial}{\partial m^2} P_i(p^2,m^2) + D[P_i(p^2,m^2)] P_i(p^2,m^2)
\end{equation}

with $D[P_i(p^2,m^2)]$ being the mass dimension of $P_i(p^2,m^2)$. Using 
relations generated from the reduction, the right hand side can again
be expressed through master integrals which leads to a coupled system
of inhomogeneous differential equations

\begin{equation}
  \frac{d}{d z} P_i(z) = A_{ij}(z,\epsilon) P_j(z). \label{eq:deqnsys}
\end{equation}

The matrix $A_{ij}(z,\epsilon)$ is composed of rational functions of
$z$ and $\epsilon$. Its block-triangular form simplifies the problem
to a set of several small coupled systems of differential equations
and therefore provides a systematic approach to solutions of $P_i(z)$
in terms of expansions in the external momentum $z$ (or $y=-z^{-1}$).

\medskip

In the low- ($z \to 0$) and high-energy ($y \to 0$) limits, the system
was solved by ans\"atze similar to eqns.~(\ref{eq:lowzanzatz},
\ref{eq:largezanzatz}) for each master integral and coefficients of
the series were determined recursively up to high powers in $z$ and
$y$, respectively. 

\begin{figure}[h]
  \begin{minipage}{0.5\textwidth}
     \begin{center}
       \epsfig{file=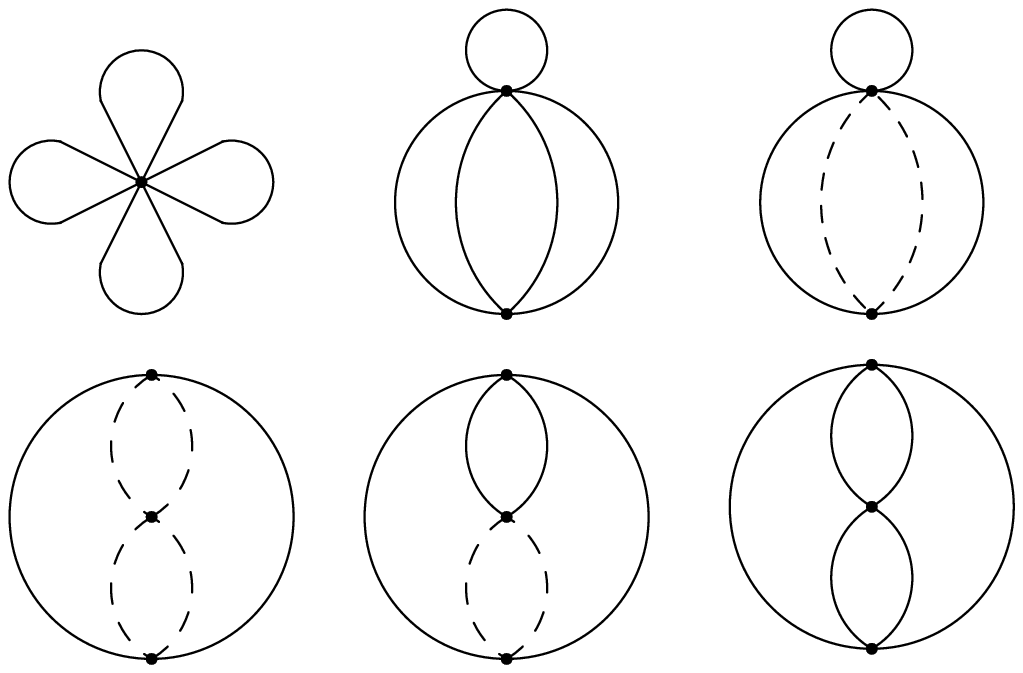, width=.8\textwidth}\\
       (a)
     \end{center}
  \end{minipage}
  \begin{minipage}{0.5\textwidth}
     \begin{center}
       \epsfig{file=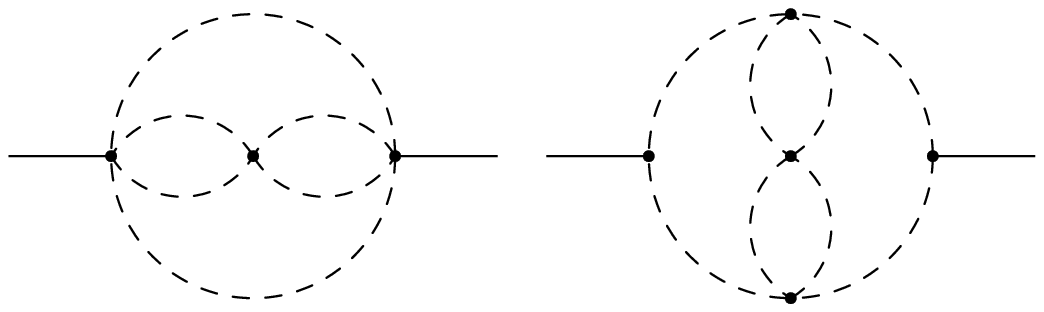, width=.8\textwidth}\\
       (b)
     \end{center}
  \end{minipage}
  \caption{Boundary conditions appearing in the expansions of master
  integrals}
  \label{fig:boundaries}
\end{figure}

Boundary conditions in the low energy limit are given by massive
tadpole diagrams depicted in fig.~\ref{fig:boundaries}(a),
analytically  calculated in~\cite{Chetyrkin:2006dh}.
In the opposite limit, all boundary conditions were determined from
automatized diagrammatic large momentum expansions which lead to
products of at most three loop massive tadpoles with massless
propagators.  Fig.~\ref{fig:boundaries}(b) illustrates the needed
propagators at the four-loop level.

The approximate linear complexity of this procedure allows for the
computation of coefficients to, at least in principle, arbitrary
depths.  In this work, we concentrate on the first 30 and 15 terms of the
expansions for $z<1$ and $y<1$, respectively, and are thus able to
compute the master integrals in those regions with high precision.

\begin{figure}
  \begin{center}
    \epsfig{file=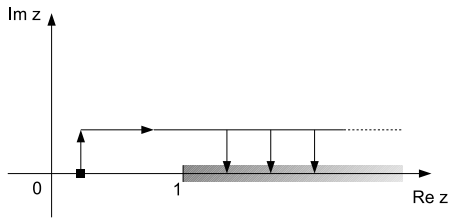, width=.4\textwidth}
  \end{center}
  \label{fig:contourDeformation}
  \caption{Integration contour chosen for the numerical integration}
\end{figure}

In addition, an exact numerical solution of eqn.~(\ref{eq:deqnsys}) was
accomplished. For this purpose, the $\epsilon$-expanded system of
differential equations was directly integrated by means of the FORTRAN
package ODEPACK~\cite{odepack}, using the high precision values at a starting
point $|z| \ll 1$. Due to the absence of thresholds in the Euclidean
domain, the numerical integration can easily be carried out along the
real axis for $z < 0$. In the Minkowskian half-plane, however,
(pseudo)thresholds occur and integration along the real axis is only
possible below these special points. On the other hand, by virtue of
contour deformation into the complex plane (see fig.
\ref{fig:contourDeformation}), the master integrals can be solved
numerically for arbitrary values of $z > 0$, even above $z=1$.
Variation of the integration contour is used to estimate the real
achieved precision.  Furthermore, the Mellin-Barnes (MB) method has
been used at a few points to check the obtained values from direct
integration. For this purpose, MB representations have been
automatically generated with the package
\texttt{MBrepresentation}~\cite{MBrep}, analytically continued in
$\epsilon$ and numerically integrated with help of the \texttt{MB}
package~\cite{Czakon:2005rk}. We observed that the high-energy expansion works
very well above at least $z>5$ and therefore the numerical integration was
undertaken up to $z=10$.  This fact is also used as a cross-check of the
high-energy expansion against the numerically computed polarization function.

In a last step, we performed the renormalization of the mass $m(\mu)$,
the strong coupling $\alpha_s(\mu)$ and the external current in the 
$\overline{\text{MS}}$-scheme. 

\section{Results}
All analytical results for the coefficients in the small- and
high-energy series in the $\overline{\text{MS}}$-scheme up to the
30th and 15th term, respectively, are available in \texttt{Mathematica}
format together with the source of this paper from
\texttt{http://arxiv.org}.  Apart from the new expressions at the four-loop
level, we also provide so far unknown terms needed for renormalization in the
large-energy expansion at the three-loop level. The Taylor series for $z\to0$
is already known to sufficient depth~\cite{Boughezal:2006uu, Maier:2007yn}.

For the sake of clarity, here we give only the first five terms of both series
at $\order(\alpha_s^3)$.  The first coefficients in the limit $z \to 0$ are
given by 
\begin{eqnarray}
 C^0_1 &=& 
   \frac{163868}{98415}
   -\frac{3287}{2430} \,\zeta_3
   +\frac{203}{324} \,l_m \zeta_3
   -\frac{14483}{21870} \,l_m
   +\frac{236}{3645} \,l_m^2
   +\frac{8}{135} \,l_m^3
   +\,n_l \left(
   \frac{262877}{262440} 
\right.\nonumber\\\nonumber&&\left.
   -\frac{116}{81} \,a_4 
   -\frac{29}{486} \,l_2^4
   +\frac{29}{486} \,l_2^2 \pi^2
   +\frac{1421}{58320} \pi ^4
   -\frac{38909}{19440} \,\zeta_3
   +\frac{203}{324} \,l_m \,\zeta_3
   -\frac{3779}{21870}\,l_m 
\right.\\\nonumber&&\left. 
   +\frac{472}{3645} \,l_m^2
   +\frac{16}{135} \,l_m^3
   \right)
   +\,n_l^2 \left(
   \frac{42173}{32805}
   -\frac{112}{135} \,\zeta_3
   +\frac{1784}{3645} \,l_m
   +\frac{236}{3645} \,l_m^2
   +\frac{8 }{135}\,l_m^3
   \right)\\
\end{eqnarray}
\begin{eqnarray}
 C^0_2 &=&
   \frac{1842464707}{646652160}
   -\frac{2744471}{1064448} \,\zeta_3
   +\frac{14203}{27648} \,l_m \,\zeta_3
   -\frac{676663}{870912} \,l_m
   -\frac{1468}{42525} \,l_m^2
   +\frac{16 }{315}\,l_m^3 
\nonumber\\\nonumber&&
   +\,n_l \left(
   \frac{95040709}{62705664}
   -\frac{2029}{1728}\,a_4
   +\frac{2029}{41472} \,l_2^2 \pi^2
   -\frac{2029}{41472} \,l_2^4
   +\frac{99421}{4976640} \pi^4
   -\frac{12159109}{4644864} \,\zeta_3
\right.\\\nonumber&&\left.
   +\frac{14203}{27648} \,l_m \,\zeta_3
   -\frac{153560977}{326592000} \,l_m
   -\frac{2936}{42525}\,l_m^2
   +\frac{32}{315} \,l_m^3
   \right) 
   +\,n_l^2 \left(
   \frac{15441973}{19136250}
   -\frac{32}{45} \,\zeta_3
\right.\\&&\left.
  +\frac{195679}{637875} \,l_m
  -\frac{1468}{42525}\,l_m^2
  +\frac{16}{315} \,l_m^3
  \right) 
\\\nonumber\\
 C^0_3 &=& 
  \frac{56877138427}{12609717120}
  -\frac{6184964549}{1556755200} \,\zeta_3
  +\frac{12355}{20736} \,l_m \,\zeta_3
  -\frac{224445289}{244944000} \,l_m
  -\frac{310736}{4465125} \,l_m^2
\nonumber\\\nonumber&&
  +\frac{128 }{2835} \,l_m^3
  +\,n_l \left(
  \frac{60361465477}{29393280000}
  -\frac{1765}{1296} \,a_4
  -\frac{1765}{31104} \,l_2^4
  +\frac{1765}{31104} \,l_2^2 \pi^2
  +\frac{17297}{746496} \pi^4
\right.\\\nonumber&&\left.
  -\frac{57669161}{17418240} \,\zeta_3
  +\frac{12355}{20736} \,l_m\,\zeta_3
  -\frac{37155186557}{60011280000}\,l_m
  -\frac{621472}{4465125} \,l_m^2
  +\frac{256}{2835} \,l_m^3
  \right)
\\&&
  +\,n_l^2 \left(
  \frac{31556642272}{49228003125}
  -\frac{256}{405} \,\zeta_3
  +\frac{15480824}{52093125} \,l_m
  -\frac{310736}{4465125}\,l_m^2
  +\frac{128}{2835} \,l_m^3
  \right)
\\\nonumber\\
 C^0_4 &=& 
  \frac{270605350139987}{40351094784000}
  -\frac{692437613459}{119558799360} \,\zeta_3
  +\frac{2522821}{3538944} \,l_m \,\zeta_3  
  -\frac{4400766606253}{4138573824000} \,l_m
\nonumber\\\nonumber&&
  -\frac{1082216}{12629925}\,l_m^2
  +\frac{256}{6237} \,l_m^3 
  +\,n_l \left(
  \frac{432564184014463}{165542952960000}
  -\frac{360403}{221184} \,a_4
  -\frac{360403}{5308416}\,l_2^4
\right.\\\nonumber&&\left.
  +\frac{360403}{5308416} \,l_2^2 \pi^2
  +\frac{17659747}{637009920} \pi^4
  -\frac{44387709491 }{10899947520}\,\zeta_3
  +\frac{2522821 }{3538944} \,l_m\,\zeta_3
  -\frac{143778598477}{189621927936} \,l_m
\right.\\\nonumber&&\left.
  -\frac{2164432}{12629925} \,l_m^2
  +\frac{512}{6237}\,l_m^3
  \right)
  +\,n_l^2\left(
  \frac{667234795424}{1253204308125}
  -\frac{512}{891} \,\zeta_3
\right.\\&&\left.
  +\frac{1213878224}{3978426375} \,l_m
  -\frac{1082216}{12629925}\,l_m^2
  +\frac{256}{6237} \,l_m^3
  \right)
\\\nonumber\\
 C^0_5 &=& 
  \frac{1089264797862809}{114328101888000}
  -\frac{13766684598973}{1693749657600} \,\zeta_3
  +\frac{1239683}{1474560} \,l_m \,\zeta_3
  -\frac{9879141778777}{8136640512000} \,l_m
\nonumber\\\nonumber&&
  -\frac{1064635136}{11345882625}\,l_m^2
  +\frac{1024}{27027} \,l_m^3
  +\,n_l \left(
  \frac{2680576384798219}{838813667328000}
  -\frac{1239683}{645120} \,a_4
  -\frac{1239683}{15482880}\,l_2^4
\right.\\\nonumber&&\left.
  +\frac{1239683}{15482880} \,l_2^2 \pi^2
  +\frac{8677781}{265420800} \pi^4
  -\frac{1942577613953}{398529331200} \,\zeta_3
  +\frac{1239683}{1474560} \,l_m \,\zeta_3
\right.\\\nonumber&&\left.
  -\frac{580030100117579693}{644112110315520000} \,l_m
  -\frac{2129270272}{11345882625} \,l_m^2
  +\frac{2048}{27027} \,l_m^3\right)
\\\nonumber&&
  +\,n_l^2 \left(
  \frac{61443753281463008}{136221219619340625}
  -\frac{2048}{3861} \,\zeta_3
  +\frac{160295587080064}{511075282843125} \,l_m
  -\frac{1064635136}{11345882625} \,l_m^2
\right.\\&&\left.
  +\frac{1024}{27027} \,l_m^3
  \right)
\end{eqnarray}
where $l_m = \log(m(\mu)^2/\mu^2)$, $l_2 = \log(2)$, $a_i =
\text{Li}_i(1/2)$ with $\text{Li}_i(x)$ the polylogarithm function and
$\zeta_i$ the Riemann zeta numbers.

\newpage

In the limit $z\to 
-\infty$ we obtain
\begin{eqnarray}
 C^\infty_0 &=&
  \frac{783389}{102060}
  -\frac{24349}{5670} \,\zeta_3
  -\frac{20}{9}\,\zeta_5 
  +\frac{7}{24} \,\zeta_3 \,l_{xm}
  -\frac{113}{324}\,l_{xm} 
  -\frac{1}{27}\,l_{xm}^2
  +\frac{1}{27}\,l_{xm}^3 
  +\frac{545}{216} \,l_{x\mu} \,\zeta_3
\nonumber\\\nonumber&&
  -\frac{365}{108} \,l_{x\mu}
  -\frac{4}{9}\,l_{x\mu}^2 \,\zeta_3
  +\frac{31}{54} \,l_{x\mu}^2
  -\frac{2}{27} \,l_{x\mu}^3
  -\frac{1}{9}\,l_{x\mu} \,l_{xm}^2
  +\frac{2}{27} \,l_{x\mu} \,l_{xm}
  +\frac{1 }{9}\,l_{x\mu}^2 \,l_{xm}
\\\nonumber&&
  +\,n_l \left(
  \frac{94735}{5832}
  -\frac{2855}{324}\,\zeta_3
  -\frac{49}{4320} \pi^4
  -\frac{40}{9} \,\zeta_5
  +\frac{2}{3} \,a_4
  +\frac{1}{36}\,l_2^4
  -\frac{1}{36}\pi^2 \,l_2^2
  +\frac{7}{24} \,l_{xm} \,\zeta_3 
\right.\\\nonumber&&\left.
  -\frac{37}{324} \,l_{xm}
  -\frac{2}{27} \,l_{xm}^2
  +\frac{2}{27} \,l_{xm}^3
  +\frac{1153}{216} \,l_{x\mu} \,\zeta_3
  -\frac{793}{108} \,l_{x\mu}
  -\frac{8}{9} \,l_{x\mu}^2\,\zeta_3
  +\frac{31}{27} \,l_{x\mu}^2
  -\frac{4}{27} \,l_{x\mu}^3
\right.\\\nonumber&&\left.
  -\frac{2}{9} \,l_{xm}^2\,l_{x\mu} 
  +\frac{4}{27} \,l_{xm} \,l_{x\mu} 
  +\frac{2}{9} \,l_{xm} \,l_{x\mu}^2\right)
  +\,n_l^2 \left(
  \frac{22327}{2916} 
  -\frac{358 }{81}\,\zeta_3
  -\frac{20}{9} \,\zeta_5 
  +\frac{19}{81} \,l_{xm} 
\right.\\\nonumber&&\left.
  -\frac{1}{27}\,l_{xm}^2 
  +\frac{1}{27}\,l_{xm}^3
  +\frac{76}{27} \,l_{x\mu} \,\zeta_3
  -\frac{107 }{27}\,l_{x\mu} 
  -\frac{4}{9} \,l_{x\mu}^2 \,\zeta_3
  +\frac{31}{54} \,l_{x\mu}^2 
  -\frac{2}{27} \,l_{x\mu}^3
  -\frac{1}{9}\,l_{x\mu} \,l_{xm}^2
\right.\\&&\left.
  +\frac{2}{27} \,l_{x\mu} \,l_{xm} 
  +\frac{1}{9}\,l_{x\mu}^2 \,l_{xm}
  \right) 
\\\nonumber\\
 C^\infty_1 &=&
  \frac{16681}{1944}
  -\frac{38}{9} \,\zeta_3 
  +\frac{8}{3} \,\zeta_3 \,l_{x\mu} 
  -\frac{317}{54}\,l_{x\mu} 
  +\frac{13}{18} \,l_{x\mu}^2
  -\frac{1}{9}\,l_{x\mu}^3
  +\,n_l \left(
  \frac{10921}{972}
  -4 \,\zeta_3 
  +\frac{8 }{3}\,\zeta_3 \,l_{x\mu} 
\right.\nonumber\\&&\left.
  -\frac{221}{27} \,l_{x\mu} 
  +\frac{13}{9} \,l_{x\mu}^2
  -\frac{2}{9}\,l_{x\mu}^3 
  \right)
  +\,n_l^2 \left( 
  \frac{5161}{1944} 
  +\frac{2}{9} \,\zeta_3 
  -\frac{125}{54}\,l_{x\mu} 
  +\frac{13}{18} \,l_{x\mu}^2 
  -\frac{1}{9}\,l_{x\mu}^3
  \right)
\\\nonumber\\
 C^\infty_2 &=& 
  -\frac{7633}{9720} 
  -\frac{571}{1620} \,\zeta_3
  + \frac{20}{9} \,\zeta_5 
  -\frac{23}{18} \,\zeta_3 \,l_{xm}
  +\frac{6373}{3888}\,l_{xm} 
  +\frac{355}{432} \,l_{xm}^2 
  +\frac{23}{216} \,l_{xm}^3 
  +\frac{1}{72}\,l_{xm}^4
\nonumber\\\nonumber&&
  +\frac{44}{27}\,\zeta_3\,l_{x\mu} 
  -\frac{1505}{1296} \,l_{x\mu} 
  +\frac{1}{9}\,\zeta_3\,l_{x\mu}^2 
  +\frac{1}{216}\,l_{x\mu}^2 
  -\frac{23}{72} \,l_{x\mu} \,l_{xm}^2 
  +\frac{1}{3}\,\zeta_3 \,l_{x\mu} \,l_{xm}
  -\frac{101}{54} \,l_{x\mu} \,l_{xm} 
\\\nonumber&& 
  +\frac{7}{24} \,l_{x\mu}^2 \,l_{xm}
  -\frac{1}{18}\,l_{x\mu} \,l_{xm}^3 
  +\frac{1}{12}\,l_{x\mu}^2 \,l_{xm}^2 
  -\frac{1}{18}\,l_{x\mu}^3 \,l_{xm}
  +\,n_l \left( 
  -\frac{12733}{7776}
  +\frac{29}{324} \,\zeta_3 
  +\frac{1}{45}\pi^4 
\right.\\\nonumber&&\left.  
  +\frac{25 }{9}\,\zeta_5
  -\frac{1}{2}\,\zeta_3 \,l_{xm} 
  +\frac{6113}{3888} \,l_{xm} 
  +\frac{455}{432} \,l_{xm}^2
  +\frac{43}{216} \,l_{xm}^3 
  +\frac{1}{36}\,l_{xm}^4 
  +\frac{16}{27}\,\zeta_3 \,l_{x\mu} 
  -\frac{397}{324} \,l_{x\mu} 
\right.\\\nonumber&&\left.  
  +\frac{2 }{9} \,\zeta_3\,l_{x\mu}^2 
  +\frac{1}{108}\,l_{x\mu}^2 
  -\frac{43}{72} \,l_{x\mu} \,l_{xm}^2 
  +\frac{1}{3}\,\zeta_3 \,l_{x\mu} \,l_{xm}
  -\frac{257}{108} \,l_{x\mu} \,l_{xm} 
  +\frac{7}{12} \,l_{x\mu}^2 \,l_{xm} 
  -\frac{1}{9}\,l_{x\mu} \,l_{xm}^3
\right.\\\nonumber&&\left.
  +\frac{1}{6}\,l_{x\mu}^2 \,l_{xm}^2 
  -\frac{1}{9}\,l_{x\mu}^3 \,l_{xm}
  \right) 
  +\,n_l^2 \left( 
  -\frac{233}{7776} 
  +\frac{239}{162} \,\zeta_3 
  -\frac{1}{90}\pi^4 
  +\frac{5}{9} \,\zeta_5
  +\frac{7}{9} \,\zeta_3 \,l_{xm} 
  +\frac{89}{486} \,l_{xm} 
\right.\\\nonumber&&\left.  
  +\frac{25}{108} \,l_{xm}^2
  +\frac{5}{54} \,l_{xm}^3 
  +\frac{1}{72}\,l_{xm}^4 
  -\frac{28}{27}\,\zeta_3 \,l_{x\mu} 
  -\frac{83}{1296} \,l_{x\mu}
  +\frac{1}{9}  \,\zeta_3\,l_{x\mu}^2 
  +\frac{1}{216}\,l_{x\mu}^2 
  -\frac{5}{18} \,l_{x\mu} \,l_{xm}^2
\right.\\&&\left.  
  -\frac{55}{108} \,l_{x\mu} \,l_{xm} 
  +\frac{7}{24} \,l_{x\mu}^2 \,l_{xm}
  -\frac{1}{18}\,l_{x\mu} \,l_{xm}^3 
  +\frac{1}{12}\,l_{x\mu}^2 \,l_{xm}^2 
  -\frac{1}{18}\,l_{x\mu}^3 \,l_{xm}
  \right) 
\\\nonumber\\
 C^\infty_3 &=& 
  -\frac{205123}{1049760} 
  -\frac{593 }{2430}\,\zeta_3 
  -\frac{5 }{54}\,\zeta_3 \,l_{xm}
  +\frac{15017 }{23328}\,l_{xm} 
  +\frac{325 }{3888}\,l_{xm}^2 
  -\frac{47 }{5832}\,l_{xm}^3
  +\frac{1}{324}\,l_{xm}^4 
\nonumber\\\nonumber&& 
  +\frac{35 }{81}\,\zeta_3\,l_{x\mu} 
  +\frac{7963 }{8748}\,l_{x\mu} 
  -\frac{227 }{1944}\,l_{x\mu}^2 
  +\frac{1}{36}\,l_{x\mu}^3
  -\frac{7 }{486}\,l_{x\mu} \,l_{xm}^3 
  +\frac{25 }{648}\,l_{x\mu}^2 \,l_{xm}^2
  -\frac{1}{36}\,l_{x\mu}^3 \,l_{xm} 
\\\nonumber&&
  -\frac{109 }{1944}\,l_{x\mu} \,l_{xm}^2 
  -\frac{1355 }{1944}\,l_{x\mu} \,l_{xm}
  +\frac{283 }{1944}\,l_{x\mu}^2 \,l_{xm} 
  +\,n_l \left( 
  \frac{49255}{104976}
  -\frac{167 }{162}\,\zeta_3 
  +\frac{7}{1215}\pi^4 
\right.\\\nonumber&&\left.
  +\frac{16 }{81}\,\zeta_3 \,l_{xm} 
  +\frac{4987 }{3888}\,l_{xm} 
  +\frac{3149 }{11664}\,l_{xm}^2
  +\frac{175 }{5832}\,l_{xm}^3 
  +\frac{7 }{972}\,l_{xm}^4 
  +\frac{23 }{81}\,l_{x\mu} \,\zeta_3
  +\frac{29663 }{34992}\,l_{x\mu} 
\right.\\\nonumber&&\left.  
  -\frac{227 }{972}\,l_{x\mu}^2
  +\frac{1}{18}\,l_{x\mu}^3 
  -\frac{5 }{162}\,l_{x\mu} \,l_{xm}^3 
  +\frac{25 }{324}\,l_{x\mu}^2 \,l_{xm}^2
  -\frac{1}{18}\,l_{x\mu}^3 \,l_{xm} 
  -\frac{367 }{1944}\,l_{x\mu} \,l_{xm}^2
\right.\\\nonumber&&\left.
  -\frac{1597 }{1458}\,l_{x\mu} \,l_{xm} 
  +\frac{283 }{972}\,l_{x\mu}^2 \,l_{xm}
  \right)
\end{eqnarray}
\begin{eqnarray}
&&
  +\,n_l^2 \left(
  \frac{9157}{209952} 
  +\frac{53 }{243}\,\zeta_3 
  -\frac{4}{1215} \pi^4 
  +\frac{59 }{162}\,\zeta_3 \,l_{xm}
  +\frac{5905 }{23328}\,l_{xm} 
  +\frac{709 }{5832}\,l_{xm}^2 
  +\frac{7 }{243}\,l_{xm}^3 
\right.\nonumber\\\nonumber&&\left.
  +\frac{1}{243}\,l_{xm}^4
  -\frac{4 }{27} \,\zeta_3\,l_{x\mu} 
  -\frac{2189 }{34992}\,l_{x\mu} 
  -\frac{227 }{1944}\,l_{x\mu}^2
  +\frac{1}{36}\,l_{x\mu}^3 
  -\frac{4 }{243}\,l_{x\mu} \,l_{xm}^3 
  +\frac{25 }{648}\,l_{x\mu}^2 \,l_{xm}^2
\right.\\&&\left.
  -\frac{1}{36}\,l_{x\mu}^3 \,l_{xm} 
  -\frac{43 }{324}\,l_{x\mu} \,l_{xm}^2
  -\frac{2323 }{5832}\,l_{x\mu} \,l_{xm} 
  +\frac{283 }{1944}\,l_{x\mu}^2 \,l_{xm}
  \right)  
\\\nonumber\\
 C^\infty_4 &=& 
  -\frac{35389447}{44789760} 
  -\frac{227 }{103680}\,\zeta_3 
  -\frac{895 }{1728}\,\zeta_3 \,l_{xm}
  +\frac{2812987 }{2985984}\,l_{xm} 
  +\frac{125203 }{248832}\,l_{xm}^2 
  +\frac{4603 }{62208}\,l_{xm}^3
\nonumber\\\nonumber&& 
  +\frac{487 }{41472}\,l_{xm}^4 
  +\frac{31 }{64}\,\zeta_3\,l_{x\mu} 
  +\frac{164053 }{165888}\,l_{x\mu} 
  -\frac{70553 }{497664}\,l_{x\mu}^2 
  +\frac{7 }{288}\,l_{x\mu}^3
  -\frac{185 }{5184}\,l_{x\mu} \,l_{xm}^3 
\\\nonumber&&
  +\frac{229 }{6912}\,l_{x\mu}^2 \,l_{xm}^2 
  -\frac{1}{48}\,l_{x\mu}^3 \,l_{xm} 
  -\frac{8501 }{41472}\,l_{x\mu} \,l_{xm}^2 
  -\frac{80177 }{124416}\,l_{x\mu} \,l_{xm}
  +\frac{2279 }{20736}\,l_{x\mu}^2 \,l_{xm} 
\\\nonumber&&
  +\,n_l \left( 
  -\frac{19473727}{35831808}
  -\frac{27725 }{20736}\,\zeta_3   
  +\frac{43}{6480} \pi^4
  +\frac{271 }{1728}\,\zeta_3 \,l_{xm} 
  +\frac{3637091 }{2985984}\,l_{xm} 
  +\frac{226123 }{497664}\,l_{xm}^2
\right.\\\nonumber&&\left.
  +\frac{11407 }{124416}\,l_{xm}^3 
  +\frac{19 }{1296}\,l_{xm}^4 
  +\frac{89 }{192}\,l_{x\mu} \,\zeta_3 
  +\frac{47029 }{41472}\,l_{x\mu} 
  -\frac{70553 }{248832}\,l_{x\mu}^2
  +\frac{7 }{144}\,l_{x\mu}^3 
\right.\\\nonumber&&\left.
  -\frac{497 }{10368}\,l_{x\mu} \,l_{xm}^3 
  +\frac{229 }{3456}\,l_{x\mu}^2 \,l_{xm}^2
  -\frac{1}{24}\,l_{x\mu}^3 \,l_{xm}   
  -\frac{1673 }{5184}\,l_{x\mu} \,l_{xm}^2
  -\frac{57629 }{62208}\,l_{x\mu} \,l_{xm} 
\right.\\\nonumber&&\left.
  +\frac{2279 }{10368}\,l_{x\mu}^2 \,l_{xm} 
  \right)
  +\,n_l^2 \left(
  -\frac{3462161}{35831808} 
  -\frac{745 }{5184}\,\zeta_3 
  -\frac{127}{51840} \pi ^4
  +\frac{265 }{864}\,\zeta_3 \,l_{xm} 
  +\frac{1097 }{5832}\,l_{xm}
\right.\\\nonumber&&\left.
  +\frac{51077 }{497664}\,l_{xm}^2 
  +\frac{2411 }{124416}\,l_{xm}^3 
  +\frac{127 }{41472}\,l_{xm}^4
  -\frac{1}{48}\,\zeta_3\,l_{x\mu} 
  +\frac{8021 }{55296}\,l_{x\mu} 
  -\frac{70553 }{497664}\,l_{x\mu}^2
\right.\\\nonumber&&\left. 
  +\frac{7 }{288}\,l_{x\mu}^3 
  -\frac{127 }{10368}\,l_{x\mu} \,l_{xm}^3
  +\frac{229 }{6912}\,l_{x\mu}^2 \,l_{xm}^2 
  -\frac{1}{48}\,l_{x\mu}^3 \,l_{xm} 
  -\frac{4883 }{41472}\,l_{x\mu} \,l_{xm}^2
\right.\\&&\left.
  -\frac{35081 }{124416}\,l_{x\mu} \,l_{xm} 
  +\frac{2279 }{20736}\,l_{x\mu}^2 \,l_{xm} 
  \right)
\end{eqnarray}
with $l_{x\mu} = \log(-p^2/\mu^2)$ and $l_{xm} = \log(-p^2/m(\mu)^2)$. 

\begin{figure}[t]
  \begin{center}
    \epsfig{file=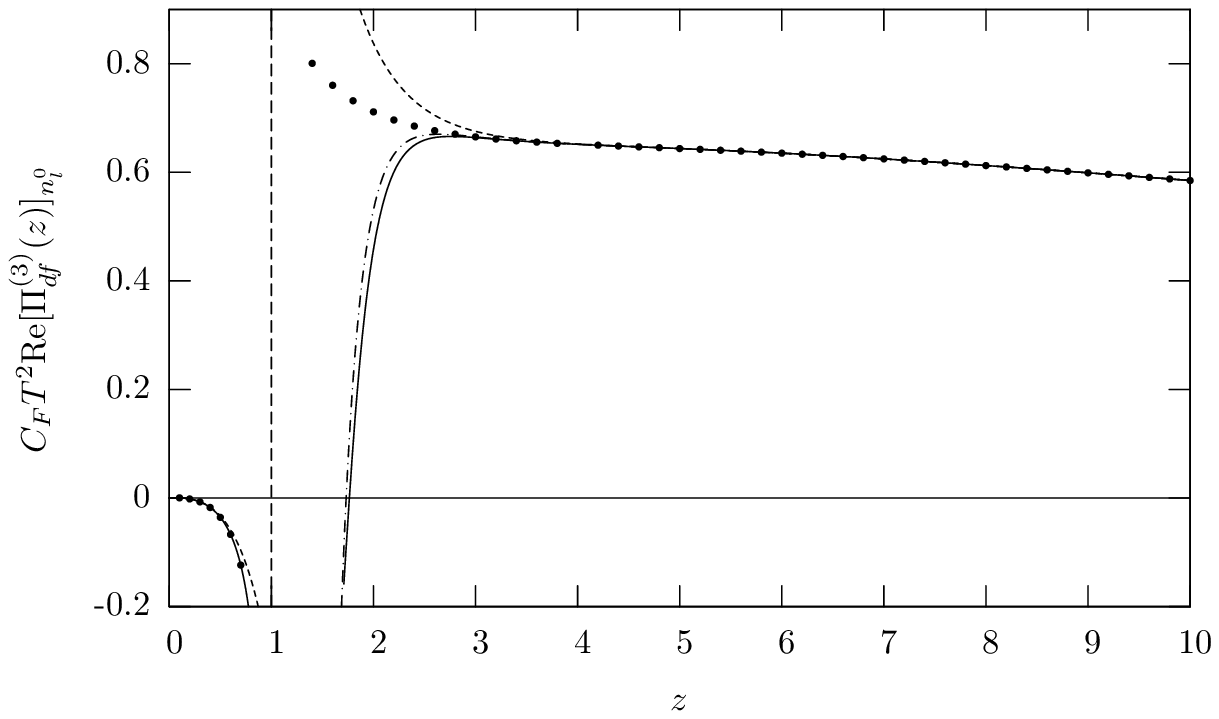, width=.75\textwidth}
    \epsfig{file=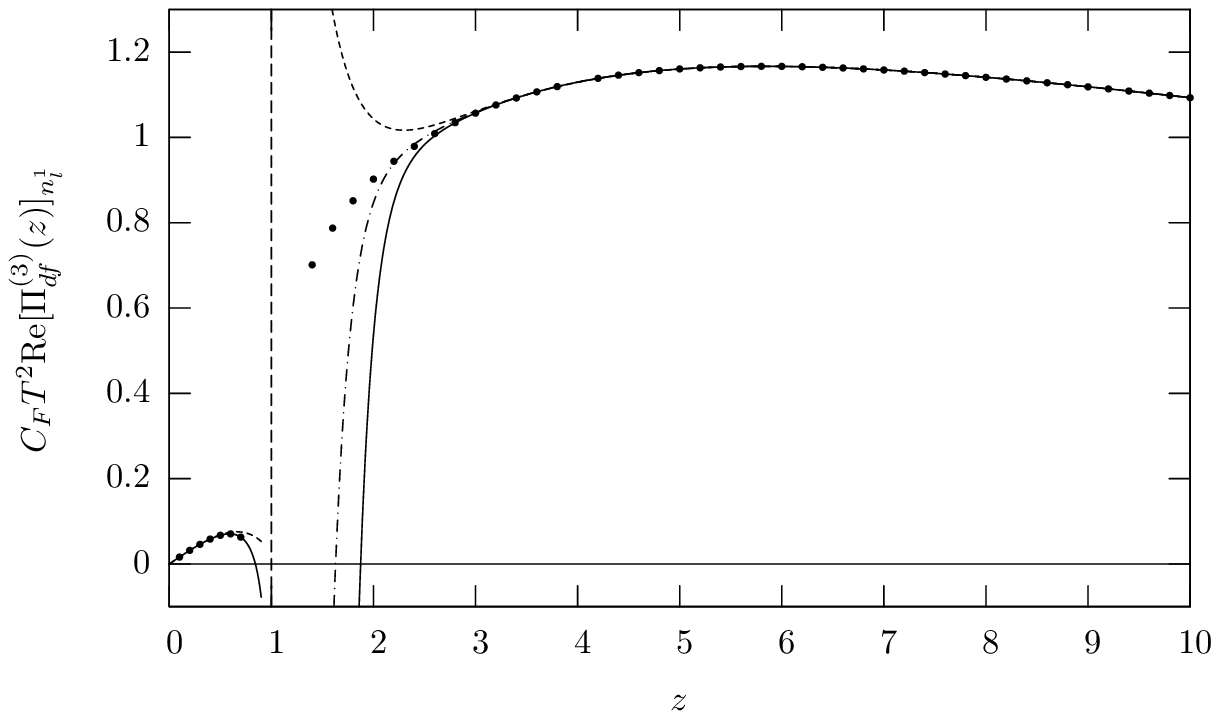, width=.75\textwidth}
    \epsfig{file=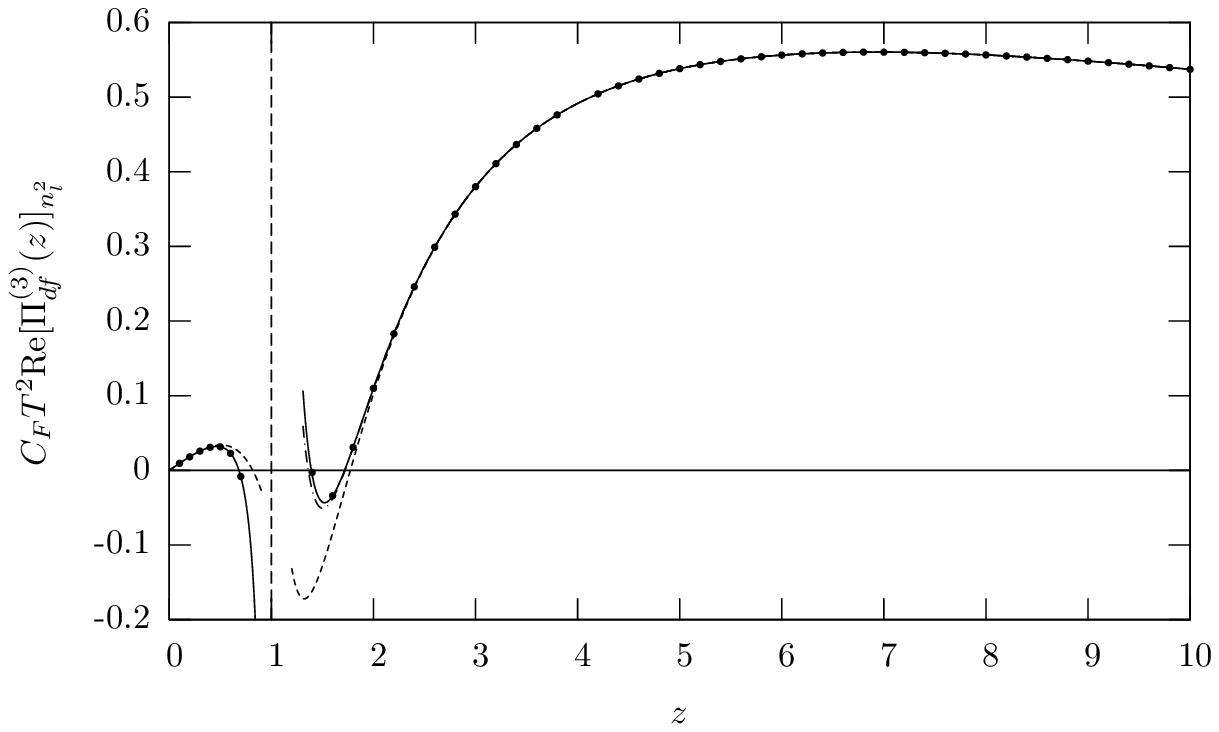, width=.75\textwidth}
    \caption{
    Comparison between expansions and numerics of the real part
    of $C_FT^2 \Pi_{\text{df}}^{(3)}(z)$ for each coefficient of $n_l$
    separately.    
    Below threshold (represented by the vertical line in $z=1$), the
    dashed (solid) curves correspond to low-energy expansions
    including the first 5 (30) terms. Above threshold
    the dashed, dash-dotted and solid lines denote the high-energy
    expansions including the first 5, 10 and 15 terms, respectively.} 
    \label{fig:plotfirst} 
  \end{center} 
\end{figure}

\begin{figure}[t]
  \begin{center}
    \epsfig{file=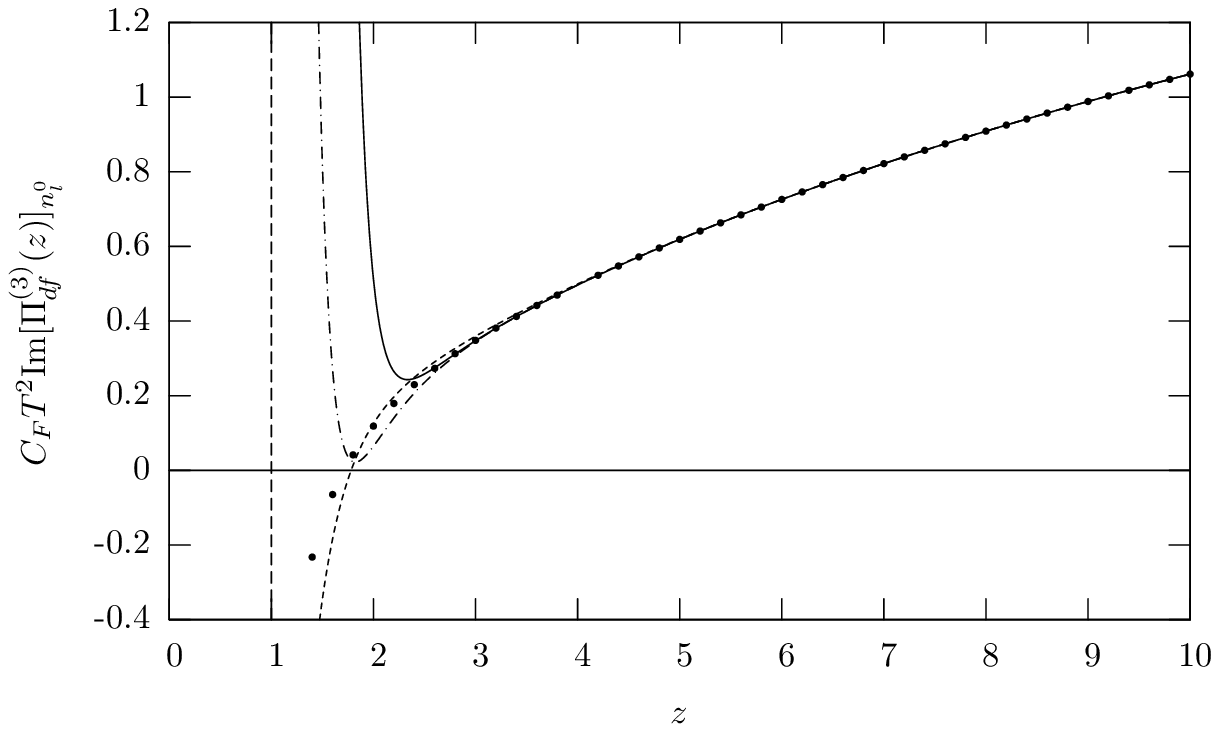, width=.75\textwidth}
    \epsfig{file=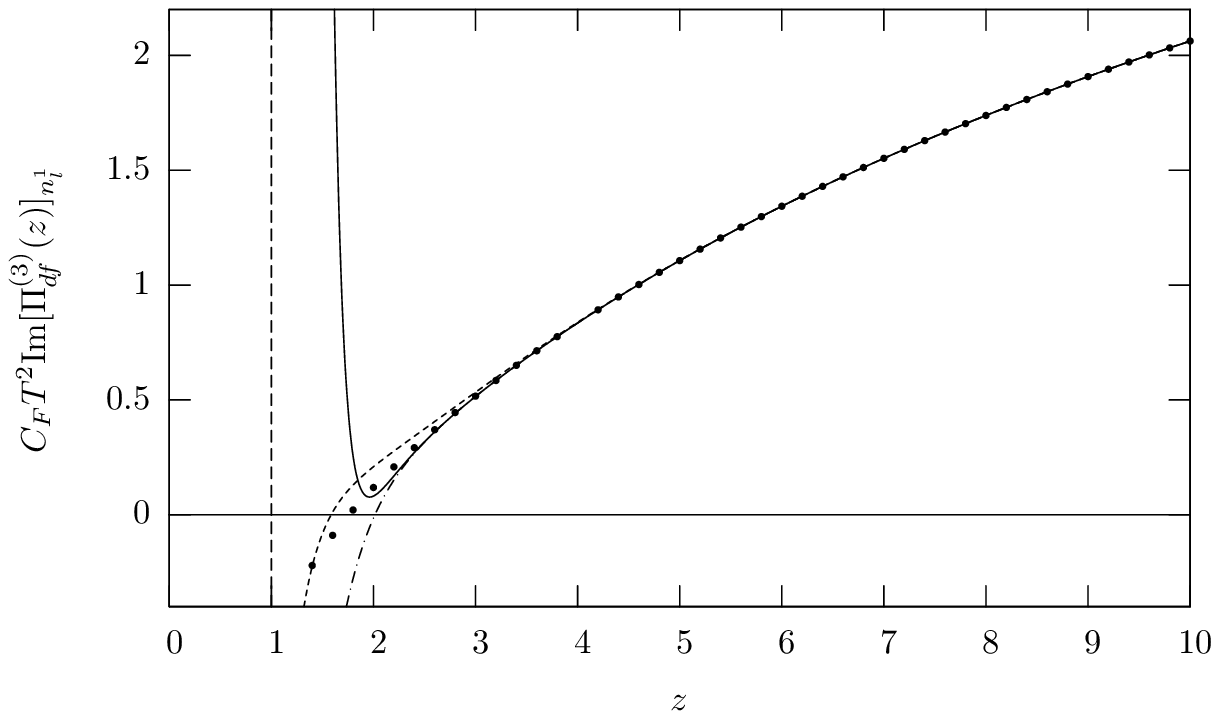, width=.75\textwidth}
    \epsfig{file=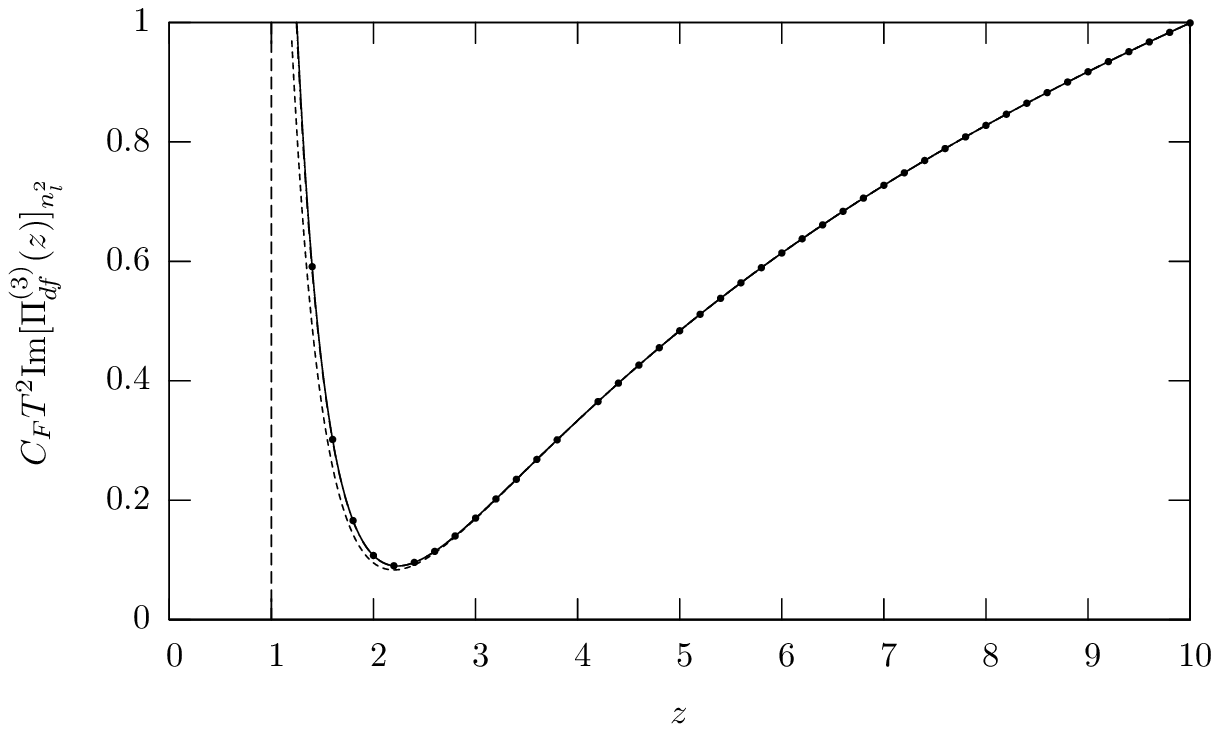, width=.75\textwidth}
    \caption{
    Comparison between expansions and numerics of the imaginary part
    of $C_FT^2 \Pi_{\text{df}}^{(3)}(z)$ for each coefficient of $n_l$
    separately. The dashed, dash-dotted and solid lines denote the
    high-energy expansions including the first 5, 10 and 15 terms,
    respectively.} 
    \label{fig:plotlast}
  \end{center}
\end{figure}

In fig. \ref{fig:plotfirst}-\ref{fig:plotlast} the exact result
obtained from numerical integration is compared against the asymptotic
behaviours of different expansion depths for Minkowskian momenta with
$m=\mu$. In both regimes, we find that already a moderate number of
terms reproduces the exact curve very well even close to
threshold.

\section{Conclusions}
In this work, the low- and high-energy expansions together with an
exact numerical solution of the double fermionic contribution to the
heavy quark vector current correlator in four-loop approximation were 
obtained. We have shown that the method of differential equations provides an
excellent tool to compute this quantity with high precision in the whole
momentum region. Consequently, the completion of the IBP reduction represents 
the only remaining task to obtain the full vacuum polarization at this order.

\newpage
\section*{Acknowledgements}
This work was supported by the Sofja Kovalevskaja Award of the
Alexander von Humboldt Foundation sponsored by the German Federal
Ministry of Education and Research.

\end{document}